\documentclass[pra,twocolumn, showpacs,nofootinbib]{revtex4}
\usepackage{amstext,amsmath,amssymb,amsfonts}
\usepackage[dvips]{graphicx}
\usepackage{latexsym}
\usepackage{euscript}
\newcommand{\Ref}[1]{(\ref{#1})}

\newcommand{\N}{\mathbb{N}}

\newcommand{\C}{\mathbb{C}}

\renewcommand{\S}{\sigma}

\DeclareMathOperator{\tr}{tr}

\def\be{\begin{equation}}
\def\ee{\end{equation}}
\def\bes{\begin{eqnarray}}
\def\ees{\end{eqnarray}}
\def\nn{\nonumber}

\newcommand{\SU}{\mathrm{SU}}
\newcommand{\U}{\mathrm{U}}

\def\hh{{\cal H}}

\def\6{\langle}
\def\9{\rangle}
\def\tr{{\rm tr}\,}
\def\half{\mbox{$1\over2$}}
\def\ha3{\mbox{$3\over2$}}
\def\bJ{{\bf J}}
\def\eS{\EuScript{S}}
\def\1{\mbox{1\hskip-.25em l}}

\begin{document}
\title{Entanglement of zero angular momentum mixtures and black hole entropy}
\author{{\bf Etera R. Livine}\footnote{elivine@perimeterinstitute.ca},
{\bf Daniel R. Terno}\footnote{dterno@perimeterinstitute.ca} }
\affiliation{Perimeter Institute, 31 Caroline St, Ontario, Canada N2L 2Y5}

\begin{abstract}

We calculate the entanglement of formation and the entanglement of
distillation for arbitrary mixtures of the zero spin states on an
arbitrary-dimensional bipartite Hilbert space.  Such states are
relevant to quantum black holes and to decoherence-free subspaces
based communication. The two measures of entanglement are equal
and scale logarithmically with the system size. We discuss its
relation to the black hole entropy law. Moreover, these states are
locally distinguishable but not locally orthogonal, thus violating
a conjecture that the entanglement measures coincide only on
locally orthogonal states. We propose a slightly weaker form of
this conjecture. Finally, we generalize our entanglement analysis
to any unitary group.
\end{abstract}
\maketitle




For slightly more than a decade entanglement is a resource of
quantum information theory   and not just a mystery in foundations
of quantum mechanics
\cite{qit,nie}. It is recognized in  systems ranging from spin chains
\cite{qrit} to black holes \cite{pt}.  However, even the entanglement
between two parties remains a slightly bewildering subject,
because not much is known beyond the two-qubit mixed states. There
are various measures of a mixed-state entanglement that reflect
different aspects of their preparation and manipulation
\cite{mesrev}. Recently there was a progress in showing equivalence between
different conjectured properties of entanglement measures and
communication channels \cite{cmp} and in estimating entanglement
of generic high-dimensional bipartite states
\cite{hlw}. However,  it is rarely possible to calculate
 entanglement of a high-dimensional mixed state, unless some
 symmetries or constraints are involved \cite{eis}.

In this article  we present a class of bipartite mixed states for
which it is possible to calculate entanglement exactly. Moreover,
we show that for this  family all standard measures of
entanglement
\cite{mesrev} are equal. We start from a parti\-cular state that
actually appears in the black hole entropy calculations
(for more details see \cite{lt}). It is the completely mixed state on the $\SU(2)$
invariant space of an arbitrary number of qubits. We then
generalize it to the arbitrary
 mixtures of zero angular momentum states in any dimension. These
 states are locally distinguishable but not locally orthogonal (in
 the sense of \cite{hor98}).  As a result, they form a
  counterexample to the conjectures of \cite{hor98}, but satisfy
  its slightly weaker form (see below).
These ideas are also relevant to decoherence-free subspaces of qubits and to
communication protocols without a shared refe\-rence frame \cite{stevie}.

In the simplest scenario, the black hole entropy calculation is
reduced to counting the number of distinct $\SU(2)$ invariant
states on the space of $2n$ qubits. This can be understood within the framework of loop quantum gravity \cite{lt}.
The ignorance of a
particular microstate makes the statistical state under
consideration to be a maximally mixed state on the zero angular
momentum $\bJ^2=0$ subspace of $(\C^2)^{\otimes 2n}$ (space of
intertwiners). We label this subspace as $\hh_0$ and the state as
$\rho$. Its orthogonal decomposition is simply \be
\rho=\frac{1}{N}\sum_i |\Psi_i\9\6\Psi_i|, \label{rho} \ee where
$|\Psi_i\9$  span a basis of $\hh_0$, $\bJ^2|\Psi_i\9=0$, and
$N\equiv{\rm dim}\,\hh_0$. A straightforward calculation \cite{lt}
gives
\be N={2n\choose n}-{2n\choose n+1}=\frac{1}{n+1}{2n\choose
n}.
\label{nas}
\ee
In the limit of large $n$, this gives back the
Bekenstein-Hawking entropy and its logarithmic correction:
\be
S\equiv-\tr\rho\log\rho=\log N \sim 2n \log2-\ha3\log n.
\label{bh}
\ee
More generally, Schur's duality \cite{good} allows
to decompose the Hilbert space of the collection of $2n$ qubits as
\be
\bigotimes^{2n}\C^2\cong\bigoplus_{j=0}^{n}\hh_j\equiv\bigoplus_{j=0}^{n}
V^j\otimes \S_{n,j},\label{schur}
\ee
where $V^j$ is the
irreducible spin-$j$ representation  of $\SU(2)$, and $\S_{n,j}$
is the irreducible representation of the permutation group
$\eS_{2n}$ that corresponds to the partition $[n+j,n-j]$ of $2n$
objects \cite{wkt}. For $n=1$, it is the simple decomposition of two qubits into irreducible representations:
$$
(V^{1/2})^{\otimes 2}=V^0\oplus V^1,
$$
where the singlet state is the antisymmetrized states, while the spin-1 space is defined as the symmetrized state space. For $n=2$, we are dealing with four qubits:
$$
(V^{1/2})^{\otimes 4}=2V^0\oplus3 V^1\oplus V^2.
$$
A basis of the singlet space is defined by the two states obtained by antisymmetrizing the qubits $(12)$ and $(34)$ then symmetrizing over the couple of pairs, or dealing with the pairs $(13)$ and $(24)$.

We introduce the multiplicity coefficients
$c^{(2n)}_j\equiv{\rm dim}\,\S_{n,j}$:
\be c^{(2n)}_j={2n\choose
n+j}-{2n\choose l+n+1} ={2n\choose n+j}\frac{2j+1}{n+j+1}.
\ee
We
recognize $N=c^{(2n)}_0$. Moreover these multiplicities satisfy
$2^{2n}=\sum_j c^{(2n)}_j (2j+1)$ and ${2n\choose n}=\sum_j
c^{(2n)}_j$ \footnote{This formula has a natural interpretation in
terms of random walks and diffusion processes~\cite{lt}.}.

We are interested to find how much entanglement is contained in
arbitrary bipartite splittings of this state. Let Alice hold
$2k\leq n$ qubits and Bob handle the rest. The Hilbert spaces of
Alice and Bob are $\hh_A\equiv(\C^2)^{\otimes {2k}}$ and
$\hh_B\equiv(\C^2)^{\otimes 2n-2k}$, respectively. Using the
Schur's duality formula for both Alice and Bob, we write
\bes
\bigotimes^{2n}\C^2&=&
\bigotimes^{2k}\C^2\otimes \bigotimes^{2n-2k}\C^2 \nn\\
&=&\bigoplus_{j_A,j_B} (V^{j_A}\otimes V^{j_B}) \otimes (\S_{k,j_A}\otimes \S_{n-k,j_B}),
\ees
with $j_{A,B}$ running up to $k,n-k$. In the following, we will
also denote $\S_{k,j}$ and $\S_{n-k,j}$ as $\S^j_{A,B}$. Singlets are
obtained when $j_A=j_B\equiv j$:
\be
\hh_0=V^0\otimes \S_{n,0}
=\bigoplus_{j=0}^k V^0_{(j)}\otimes (\S_{k,j}\otimes \S_{n-k,j}),
\label{schur2}
\ee
where $V^0_{(j)}$ is the singlet state in $V^{j}\otimes V^{j}$.  Hence the dimensionality
of $\hh_0$ is related to the multiplicities $c^{A,B}_j$ through
\be
N=c^{(2n)}_0=\sum_{j=0}^{k}c_j^{(2k)} c_j^{(2n-2k)}.
\ee

The basis states of Alice and Bob are respectively labeled as
$|j,m, a_j\9$ and $|j,m, b_j\9$. Here $0\leq j\leq k (\leq n-k)$
and $-j\leq m\leq j$ have their usual meaning and the degeneracy
labels, $a_j$ and $b_j$, enumerate the different subspaces $V^j$
(i.e. count the basis vectors in the representations of $\S_{k,j}$
and $\S_{n-k,j}$). The constraint $\bJ^2=0$ ensures  the states
$|\Psi_i\9$ are the singlets on $V^j_A\otimes V^j_B$ subspaces for
$j=0,..,k,$
\be
|j,a_j,b_j\9\equiv\frac{1}{\sqrt{2j+1}}\sum_{m=-j}^j
(-1)^{j-m}|j,-m,a_j\9\otimes|j,m,b_j\9.
\ee
Following the decomposition \Ref{schur2} of $\hh_0$ a more
transparent notation for the states $|\Psi_i\9$ is
\be
|j,a_j,b_j\9\equiv|j\9_{AB}\otimes|a_j\9_{\S^j_A}\otimes|b_j\9_{\S^j_B},
\ee
where $|j\9_{AB}$ is the singlet state on $V^{j}_A\otimes
V^{j}_B$. The totally mixed state then reads
\be
\rho=
\frac{1}{N}\sum_j \sum_{a_j,b_j}|j\9\6
j|_{AB}\otimes|a_j\9\6 a_j|_{\S^j_A}\otimes|b_j\9\6 b_j|_{\S^j_B}.
\label{rho1}
\ee
Alice's reduced density matrices result from the tracing  over
$\hh_B$,
 \be
\rho_{j,a_j}=\tr_B |j,a_j,b_j\9\6 j,a_j,b_j|.
 \ee
These matrices are independent of $b_j$. There are exactly $c^B_j$
of the matrices $\rho_{j,a_j}$ and they are diagonal, because in
the angular momentum basis  the states $|j,a_j,b_j\9$ are given as
the bi-orthogonal (Schmidt) decompositions
\cite{qit}.
Moreover, the only nonzero part of $\rho_{j,a_j}$ is a sequence of
$2j+1$ terms $1/(2j+1)$,
\be
\rho_{j,a_j}=\frac{1}{2j+1}\1_{V^j}\otimes|a_j\9\6a_j|_{\S^j_A}.
\ee
 For convenience we introduce $\rho_j=\1_{V^j}/(2j+1)$. Two   matrices $\rho_{j,a_j}$
 and $\rho_{l,a_l}$ have orthogonal  supports if one or both of their indices
are different. The reduced density matrix $\rho_A=\tr_B \rho$ is
\be
\rho_A=\frac{1}{N}\sum_j \sum_{a_j}c_j^B \rho_j\otimes|a_j\9\6a_j|_{\S^j_A}.
\label{rhoA}
\ee

We consider  the entanglements of formation and  distillation.
Together they bound all entanglement measures \cite{mesrev},
\be
E_D(\rho)\leq E(\rho)\leq E_F(\rho).
\ee
The entanglement of formation is defined as follows. A state $\rho$ can be
decomposed as a convex combination of pure states,
\be
\rho=\sum_\alpha w_\alpha |\Psi_\alpha\9\6\Psi_\alpha|, \qquad \sum_\alpha w_\alpha=1,
\qquad \forall w_\alpha>0. \label{decompos}
\ee
The entanglement of formation is the averaged degree of
entanglement of the pure states $|\Psi_\alpha\9$ (the von Neumann
entropy of their reduced density matrices) minimized over all
possible decompositions
\be
E_F(\rho)=\inf_{ \{ \Psi_\alpha\}}\sum_\alpha w_\alpha
S(\rho_\alpha).
\ee
The distillable entanglement  is the ratio of the number of
(standard) maximally entangled output states $|\Phi^+\9$ over the
needed input states $\rho$, maximized over all local operations
assisted by a classical communication (LOCC), when  the limit of infinitely many inputs is taken. Hence
\be
E_D(\rho)=\sup_{LOCC}\frac{n^{\rm out}_{|\Psi^+\9}}{n^{\rm
in}_\rho}.
\ee

\noindent \textit{Lemma:}
The entanglement of formation of the state $\rho$ is exactly given by the average of the degrees of entanglement in  the $(j,a_j,b_j)$ decomposition. As the von Neumann entropy of the reduced density matrices $\rho_{j}$ is
$S(\rho_{j, a_j})=S(\rho_{j})=\log(2j+1)$, the entanglement is:
\be
E_F(\rho)=S_E\equiv\frac{1}{N}\sum_{j,a_j,b_j}S(\rho_{j,a_j})
=\frac{1}{N}\sum_{j=0}^{k}c_j^Ac_j^B\log(2j+1).
\label{ent1}
\ee

It is instructive to prove this lemma by a direct calculation. All
pure states $|\Psi_\alpha\9$ that appear in alternative
decompositions of $\rho$ 
ought to be some linear combinations of the states $|j,a_j,b_j\9$,
\be
|\Psi_\alpha\9=\sum_{j, a_j, b_j}
c_{\alpha,ja_jb_j}|j\9_{AB}\otimes|a_j\9_{\S^j_A}\otimes|b_j\9_{\S^j_B}.
\ee

The diagonal form of $\rho$  forces the coefficients
$c_{\alpha,ja_jb_j}$ to satisfy the normalization condition
\be
\sum_\alpha
w_\alpha
c_{\alpha,ja_jb_j}c^*_{\alpha,la_lb_l}=\frac{1}{N}\delta_{jl}\delta_{a_l
a_j}\delta_{b_j b_l}. \label{wnorm}
\ee
Alice's  reduced density matrices
$\rho_A(\alpha)=\tr_B|\Psi_\alpha\9\6\Psi_\alpha|$  are
\be
\rho_A(\alpha)=\sum_j\sum_{b_j}\sum_{a_j,a'_j}c_{\alpha,ja_jb_j}c^*_{\alpha,ja'_jb_j}\rho_j\otimes|a_j\9\6a'_j|_{\S^j_A}.
\ee
Introducing
\be
\lambda_{\alpha,ja_ja'_j}=\sum_{b_j}c_{\alpha,ja_jb_j}c^*_{\alpha,ja'_jb_j},
\quad \pi_j(\alpha)=\sum_{a_j} \lambda_{\alpha, j, a_j a_j},
\ee
we rewrite the reduced density matrix of
$|\Psi_\alpha\9\6\Psi_\alpha|$ as
\be
\rho_A(\alpha)=\sum_j \pi_j(\alpha) \rho_j\otimes \Lambda_j(\alpha),
\ee
where the fictitious density matrix $\Lambda$ was introduced on
the degeneracy subspace $\S^j_A$,
\be
\Lambda_j(\alpha)=\frac{1}{\pi_j(\alpha)}\sum_{a_ja'_j}\lambda_{\alpha,ja_ja'_j}|a_j\9\6a'_j|_{\S^j_A}.
\label{lambda}
\ee
From the orthogonality  of the matrices $\rho_{j, a_j}$ and
Eq.~(\ref{wnorm})it is easy to see that
\be
\sum_\alpha w_\alpha\lambda_{\alpha,ja_ja'_j}=\frac{1}{N}c_j^B\delta_{a_j
a'_j}.
\ee

The weighted average of the entanglement of the decomposition
$\{\Phi_\alpha\}$ is $
\6 S(\{\Phi_\alpha\})\9\equiv\sum_\alpha w_\alpha S(\rho_A(\alpha))$.
From Eq.~(\ref{lambda}) and the concavity property of entropy
\cite{qit,wehrl} it follows that
\begin{widetext}
\be
\6 S(\{\Phi_\alpha\})\9\geq \sum_{\alpha, j}w_\alpha
\pi_j(\alpha)\left[S(\rho_j)+S(\Lambda_j(\alpha))\right]\geq\sum_{\alpha, j}w_\alpha
\pi_j(\alpha) S(\rho_j)=\frac{1}{N}\sum_j c^A_j c^B_j S(\rho_j).
\label{qed} \ee
\end{widetext}
QED


We illustrate this result by two particularly interesting cases
\cite{lt}. For simplicity we work in the large $n$ limit.
To start with, let Alice and Bob have $n$ qubits each. In this case,
$c_j^{A,B}\equiv c_j$, so
\be
E_F(\rho|n:n)=\frac{1}{N}\sum_{j=0}^{n/2}c_j^2\log(2j+1).
\ee
At leading order
\be
E_F(\rho|n:n)\approx\log(2j_{\max}^{(n)}+1), \label{approA}
\ee
where the coefficients $c_{j}$ reach a maximal value for
$j\equiv j_{\max}^{(n)}\approx\half\sqrt{n+2}-1$. Thus,
\be
E_F(\rho|n:n)\approx\half\log n. \label{slog}
\ee
Such entanglement is at the origin of the logarithmic corrections to the black hole entropy law.
More precisely, in a model where the black hole horizon would be constructed out of independent uncorrelated qubits,
the entropy would scale simply linearly to the number of qubits $2n$. However, the requirement of invariance under $\SU(2)$ creates correlations between the horizon qubits, which are revealed through the logarithmic correction $3/2\log n$ to the entropy law formula \Ref{bh}. Here we notice a factor 3 between the actual correlation and the entanglement, which quantifies the quantum correlations. The difference quantifies the purely classical correlations, which turn out to be roughly twice the quantum correlations. It seems that this is related to a factor 3 noticed between the classical channel capacity and the quantum channel capacity for secure communication protocols based on decoherence-free subspaces \cite{factor3}.

More generally, we find that for all
bipartite splittings of the spin network with sufficiently large
number of qubits comprising the smaller space, it is possible to
show  that
\be
I_\rho(A:B)\equiv S(\rho_A)+S(\rho_B)-S(\rho)\simeq 3S_E(\rho|A:B)
\ee
(the quantum mutual information \cite{ca} between the black hole
horizon and its parts is three times the entanglement between the
halves). Moreover, if the ratio between the number of qubits is
keep fixed while $n$ is arbitrary, the logarithmic correction
$\ha3\log n$ asymptotically equals to $I_\rho(A:B)$, so the
deviation of the black hole entropy from its classical value
equals to the total amount of correlations between the halves of
spin networks that describe it (see
\cite{lt} for more details).

Now we can push the use of entanglement further. We consider the set-up when Alice
has 2 qubits and Bob holds the rest. In this case $c_{0,1}^A=1$,
and Bob's multiplicities asymptotically satisfy
 $c_0^B/N\approx1/4$, $c_1^B/N\approx 3/4$.
Hence
\be
E_F(\rho |2:2n-2)=\frac{c_1^B}{N}\log3\approx\frac{3}{4}\log 3.
\ee
The fact that this actual entanglement is not the maximal entanglement $\log4$ can be interpreted as the pair of qubits not being maximally attached to the horizon, i.e. having a non-vanishing probability of getting detached from the horizon and evaporating from the black hole. This logic can be made more precise in the context of loop quantum gravity \cite{lqg}, and the unentangled fraction $c_0^B/N$ can be shown to be related to the black hole evaporation rate \cite{lt}. An important detail is that the unentangled fraction converges to a non-zero constant value for large black holes, which means that the pair of qubits are a constant probability of evaporating independently of the size of the black hole. This seems at first in contradiction with the behavior of the Hawking radiation. Nevertheless, to turn a probability into a rate, we need a time scale. Choosing a time scale proportional to the mass to the black hole in the semi-classical regime, we recover the right expected behavior \cite{lt}. Such a time scale can be defined for example as the period of the smallest circular orbit around the black hole.


\medskip

Let us now compute the distillable entanglement of $\rho$. From
Eq.~(\ref{rho1}) it is obvious that Alice and Bob can exactly
identify the states $|j,a_j,b_j\9$ by LOCC. It suggests a simple
distillation strategy: Alice and Bob project onto one of the
states $|j,a_j,b_j\9.$  From Eq.~(\ref{rho1}) it follows that with
a probability
\be
p_j=c^A_j c^B_j/N
\ee
they end up with a singlet on $V^j_A\otimes V^j_B$. Such a state
contains $\log(2j+1)$ ebits of entanglement (it can be converted
into this number of the standard singlets). Hence the protocol
yields $S_E$ of the distillable entanglement. Since $E_D(\rho)\leq
E_F(\rho)\leq S_E$ this protocol provides a concise proof of the
Lemma and shows that
\be
E_D(\rho)= E_F(\rho)=S_E. \label{equal}
\ee

The are two natural ways in which this setup can be generalized.
First, the elementary systems that Alice and Bob hold can be of
any fixed dimension. In the case of $2n$ objects of some higher
spin $s\in\N/2$, the degeneracy subspaces $\S^j$ are not
irreducible representations of the permutation group, but the
labeling of the states  as $|j,a_j,b_j\9$ and the entire chain of
reasoning that led to Eq.~(\ref{equal}) remain unchanged. It is
interesting to note that the logarithmic dependence of the
entanglement  on the system size also persists for arbitrary
spins. Indeed, numerical simulations and explicit calculations of
the black hole entropy \cite{lt} shows that the asymptotic
estimate of Eq.~(\ref{approA}) is independent of the basic unit:
$j^{(n)}_{ \max}$ still goes proportionally to $\sqrt{n}$ at first
order, so that  $E_F(\rho|n:n)\sim\half\log n$. In particular, we
derive the same $-3/2$ factor in front of the logarithmic
correction to the black hole entropy law \Ref{bh} whatever spin
$s$ we use to construct the model of quantum black hole.

The second point is that the mixture does not have to be equally
weighted. The same steps as above lead to the following result.

\noindent \textit{Proposition:}
Consider an arbitrary convex combination of the zero spin states,
\bes
\rho= & 
\sum_j \sum_{a_j,b_j}w^{(j)}_{a_j,b_j}|j\9\6
j|_{AB}\otimes|a_j\9\6 a_j|_{\S^A}\otimes|b_j\9\6 b_j|_{\S^B},\nn
\\
 & \sum_{a_j,b_j}w^{(j)}_{a_j,b_j}=1
\label{rhoG}
\ees
all measures of entanglement for the state $\rho$ are equal to
\be
S_E(\rho)=\sum_{j,a_j,b_j} w^{(j)}_{a_j,b_j}\log(2j+1).
\label{entG}
\ee
The only change in the calculation of $E_D$ is that
\be
p_j=\sum_{a_j,b_j} w^{(j)}_{a_j,b_j},
\ee
so for any state of the form (\ref{rhoG}) its distillable
entanglement is given by Eq.~(\ref{entG}) and is equal to its
entanglement of formation.

In \cite{hor98} it was proven that  mixed states that are
represented by an ensemble $\rho=\sum_i |\Psi_i\9 \6\Psi_i|/N$ of
locally orthogonal ensembles satisfy $E_F(\rho)=E_D(\rho)$. The
family of bipartite pure states $\{ \Psi_1,\ldots\Psi_N \}$ is
locally orthogonal if it can be ordered in such a way that for
each $i$ the local reductions  satisfy
\be
\tr(\rho_X^i\rho_X^j)=0,\label{locort}
\ee
on a fixed subspace $X$, where $j>i$, and $X$ stands for either
$A$ or $B$ subspace. It was also conjectured that  the
entanglement measures coincide only for such mixed states. While
the states $\rho_{j,a_j}$ indeed satisfy Eq.~(\ref{locort}), the
reductions of the sates $|j,a_j,b_j\9$ that differ only by $b_j$
lead to the same $\rho_{j,a_j}$ and hence are identical.
Nevertheless, the states $|j,a_j,b_j\9$ are distinguishable by
LOCC, and this is enough to establish the equality of the
entanglement measures. Hence, we propose that the conjecture
should be modified, i.e. the decomposing ensemble should consist
of the states that are distinguishable by LOCC without being
destroyed.

Let us conclude with the fact that our results hold for any
unitary group. Indeed, using $\U(M)$ or $\SU(M)$ instead of
$\SU(2)$, Schur's duality and the decomposition of tensor powers
of the fundamental representation in terms of representations of the permutation group still hold.
Therefore one is lead to the same result that all measures of entanglement on arbitrary mixtures
on the invariant subspace of the tensor products of qudits are equal, with the same
consequences in terms of the black hole entropy law.


\medskip

We thank Aram Harrow, Pawel Horedecki, Martin Plenio, and Karol
\.{Z}yczkowski for discussions and helpful comments.


\begin{thebibliography}{99}
\bibitem{qit} A. Peres, {\it Quantum Theory: Concepts and Methods\/}
(Kluwer, Dordrecht, 1993).
\bibitem{nie}  M. A. Nielsen and I. L. Chuang, {\it
Quantum Computation and Quantum Information\/}  (Cambridge
University Press, New York, 2000).
\bibitem{qrit} G. Vidal, J. I. Latorre, E. Rico, and A. Kitaev,
\prl {\bf 90}, 227902 (2003).
\bibitem{pt} A. Peres and D. R. Terno, \rmp {\bf 76}, 93 (2004).
\bibitem{mesrev} M. Horodecki, Quantum Info. Comp. {\bf 1}, 3
(2001); P. Horodecki and R. Horodecki, Quantum Info. Comp. {\bf
1}, 45 (2001); D. Bru\ss, J. Math. Phys. {\bf 43}, 4237 (2002).
\bibitem{cmp} P. W. Shor,  Commun. Math. Phys. {\bf 246}, 453 (2004); K. M. R.
Audenaert and S. L. Braunstein, Commun. Math. Phys. {\bf 246}, 443
(2004); A. S. Holevo andM. E. Shirokov, Commun. Math. Phys. {\bf
249}, 417 (2004).
\bibitem{hlw} P. Hayden, D. W. Leung, and A. Winter, e-print
quant-ph/0407094 (2004).
\bibitem{eis} J. Eisert, T. Felbinger, P. Papadopoulos, M. B.
Plenio, and M. Wilkens, \prl {\bf 84}, 1611 (2000).
\bibitem{lt} E. R. Livine and D. R. Terno, in preparation (2005).
\bibitem{hor98} P. Horodecki, R. Horodecki and M. Horodecki,
Acta Phys. Slov. {\bf 48}, 141 (1998), e-print quant-ph/9805072.
\bibitem{stevie}
S. D. Bartlett and D. R. Terno, Phys. Rev. A {\bf 71}, 012302
(2005); S. D. Bartlett, T. Rudolph, and R. W. Spekkens, Phys. Rev.
Lett. {\bf 91}, 027901 (2003).
\bibitem{good} R. Goodman and N. R. Wallach, {\em Representations
and Invariants of the Clssical Groups\/} (Cambridge University
Press, Cambridge, 1998).
\bibitem{wkt} W.-K. Tung,   {\em Group Theory in Physics\/} (World Scientific, Singapore, 1985)
\bibitem{wehrl} A. Wehrl, Rev. Mod. Phys. {\bf 50}, 221 (1978).
\bibitem{lqg}
C. Rovelli, {\em Loop Quantum Gravity}, (Cambridge University
Press, Cambridge, 2004); T. Thiemann, {\em Lectures on Loop
Quantum Gravity}, Lect. Notes Phys. {\bf 631},  41 (2003),
gr-qc/0210094; A. Ashtekar, J. Lewandowski,  Class. Quant. Grav.
{\bf 21}, R53 (2004).

\bibitem{factor3}
S. D. Bartlett, T. Rudolph and R. W. Spekkens,
Phys. Rev. A {\bf 70}, 032307 (2004).
\bibitem{ca} C. Adami N. J. Cerf,  Phys. Rev. A {\bf 56}, 3470 (1997); B.
Groisman, S. Popescu, and A. Winter, e-print quant-ph/0410091.

\end{thebibliography}
\end{document}